\documentstyle[prb,aps,epsbox,epsf]{revtex}

\newtheorem{th}{Theorem}

\begin{document} 
\draft

\title{Quantum Interleaver: Quantum Error Correction for Burst Error}
\author{Shiro Kawabata\thanks {E-mail: shiro@etl.go.jp}}
\address{
Physical Science Division, Electrotechnical Laboratory, 1-1-4 Umezono, Tsukuba, 
Ibaraki 305-8568, Japan
}
%
\date{\today}
\maketitle
\begin{abstract} 

The concept of a quantum interleaver and a simple method of quantum burst-error correction are proposed.
By using the quantum interleaver, any quantum burst-errors that have occurred are spread over the interleaved code word, so that we can construct good quantum burst-error correcting codes without increasing the redundancy of the code.
We also discuss the general method  of constructing  the quantum circuit for the quantum interleaver and the quantum network. \\

\end{abstract}
%
%
%
%
%
%
%
%
%

Quantum computers are of interest because algorithms have been discovered that enable a significant speed-up over any classical algorithm~\cite{rf:QC_review1,rf:QC_review2,rf:QC_review3}.
However, they are likely to be inherently much noisier than classical computers.
One method to deal with noise and decoherence~\cite{rf:Zurek} in quantum computers is to encode the data using a quantum error-correcting code~\cite{rf:Shor,rf:QEC1,rf:QEC2}.
The key concept of the quantum error correction is to embed quantum information represented by $k$ qubits into a larger Hilbert space of $n (>k)$ qubits.
Previously, many quantum error-correcting codes have been discovered  and various theories on the quantum error-correcting codes have also been developed, e.g., Calderbank-Shor-Steane (CSS) coding~\cite{rf:CS,rf:Steane} and stabilizer coding~\cite{rf:Gottesman,rf:Calderbank,rf:GF4}.

One important assumption in the theory of quantum error-correction is that qubits in the code word are randomly disturbed when transmitting or storing the states of an $n$ qubit register.
As in a classical computer~\cite{rf:CEC}, however, errors in quantum computers will not necessarily occur independently~\cite{rf:burst1,rf:burst2,rf:burst3}. 
Although we may use the well-known quantum $random$-error-correcting codes to protect qubits from such correlated or $burst$ errors, we must increase the redundancy of the code word, i.e.,  $(n-k)/n$, with increasing the length of the burst.
On the other hand,  if details of the decoherence mechanism of qubits are known, then it might be possible to design a more efficient error-correcting method than the conventional scheme.
In this letter, we propose a method for constructing quantum burst-error-correcting codes from known quantum error-correcting codes.
This method is based on the classical $interleaving$ technique~\cite{rf:CEC}.
We shall demonstrate that we can construct good burst-error-correcting codes $without$ $increasing$ $the$ $redundancy$ of the code word by using the quantum version of the interleaver, i.e., a $quantum$ $interleaver$. 
The construction of these codes is simple and the decoding procedure is straightforward.

First let us summarize the concept of quantum error-correcting codes~\cite{rf:Ekert}.
Suppose we need to store or communicate an unknown quantum state $\left| \phi \right>$ of a quantum system in the presence of error.
Basic types of error on a single quantum state are the bit error exchanging the states $\left| 0 \right>$ and $\left| 1 \right>$, the phase error changing the relative phase of $\left| 0 \right>$ and $\left| 1 \right>$ by $\pi$, and their combination, $i.e.,$ the bit and phase error.
The bit error corresponds to the Pauli matrix $\sigma_x(\equiv X$), the phase error to $\sigma_z(\equiv Z)$ and the bit and phase error to $- i \sigma_y(=XZ)$.
Errors operating on $n$ qubit systems are represented by tensor products of the Pauli matrices and the identity operator $I$.
Bit errors in a block of $n$ qubits are defined as a sequence of $X$ transformations performed on qubits at locations specified by a binary $n$-tuple $\alpha$.
Thus, for the basis $\left| \upsilon_1\right>, \cdots, \left| \upsilon_{2^n}\right>$  of the $2^n$-dimensional Hilbert space of $n$ qubits, the bit errors can be written as
\begin{equation}
X_{\alpha} \left| \upsilon_i\right>=\left| \upsilon_i + \alpha \right>
.
\label{eqn:e1}
\end{equation}
Analogously, phase errors can be written as 
\begin{equation}
Z_{\beta} \left| \upsilon_i\right>= \left( -1 \right)^{\upsilon_i \cdot \beta} \left| \upsilon_i \right>
,
\label{eqn:e2}
\end{equation}
where the binary vector $\beta$ represents the locations of errors.
Ekert and Macchiavello~\cite{rf:Ekert} showed that due to the interaction between $t$ qubits and the environment $\left| e \right>$, the state $\left| \phi \right>$ undergoes the following entanglement:
\begin{eqnarray}
\left| \phi \right> \otimes \left| e \right>
\to
U_{qe} \left| \phi  \right> \otimes \left| e \right>
=
\sum_{\alpha}
\sum_{\beta}
X_{\alpha}
Z_{\beta}
\left| \phi \right> 
\otimes 
\left| e_{\alpha,\beta} \right>
,
\label{eqn:e3}
\end{eqnarray}
where wt(supp$[\alpha] $ $\cup$ supp$[\beta]$)$\le t$,  and  in order to correct up to $t$ errors we need only the bit and phase errors correction codes.
In this equation, $\left| e_{\alpha,\beta} \right>$ is the state of the environment which depends on $\alpha$ and $\beta$, not on state $\left| \phi' \right>$.

In the field of $classical$ coding theory, much attention has been given to the correction of burst errors~\cite{rf:CEC}.
A burst of length $l$ is defined as a vector whose only nonzero components are among $l$ successive components, the first and last of which are nonzero.
A classical binary $[n,k]$ code $C$ is said to have burst-error-correcting ability $b$, if all bursts of length $l \le b$ are correctable.
The simplest method of constructing the burst-error-correcting code is the $interleaving$ method~\cite{rf:CEC}.
By using this method, we can easily obtain good burst-error-correcting codes without increasing the redundancy.
The interleaving of $m$ code words ($C_{1}, C_{2}, \cdots, C_{m} $) is achieved  by changing the locations of bits in code words, i.e., by constructing the new code as a vector with length $nm$ as follows:
\begin{eqnarray}
\left[ C_{1}, C_{2}, \cdots, C_{m} \right]
&=&
\left[ 
           \left( C_{1,1}, C_{1,2}, \cdots, C_{1,n} \right)
		   , 
		  \left( C_{2,1}, C_{2,2}, \cdots, C_{2,n}\right)
		   , 
		   \cdots
		   ,
		   \left( C_{m,1}, C_{m,2}, \cdots, C_{m,n}\right)
\right]
\nonumber\\
&\to&
\left[ 
           \left(C_{1,1}, C_{2,1}, \cdots, C_{m,1}\right)
		   , 
		   \left(C_{1,2}, C_{2,2}, \cdots, C_{m,2}\right)
		   ,
		   \cdots
		   ,
		   \left(C_{1,n}, C_{2,n}, \cdots, C_{m,n} \right)
\right]
.
\label{eqn:e4}
\end{eqnarray}
This procedure is  equivalent to constructing the code as a two-dimensional $n\times m$ array of the form (see Fig. 1):
\begin{eqnarray}
\left[
\matrix{
    C_{1,1}  & C_{1,2}& \cdots & C_{1,n} 
    \cr
    C_{2,1}  & C_{2,2}  & \cdots  & C_{2,n} 
	\cr
    \vdots & \vdots &   & \vdots 
	\cr
     C_{m,1}   & C_{m,2}   & \cdots  & C_{m,n} 
}
\right]
.
\label{eqn:e5}
\end{eqnarray}
Every row of this array is a code word in the original $[n,k]$ code ($C_{1}, C_{2}, \cdots, C_{m} $).
A burst of length $bm$ or less can have, at most, $b$ symbols in any row of the array in eq.~(\ref{eqn:e5}) (see also Fig. 1).
Since each row can correct a burst of length $b$ or less, the code can correct all bursts of length $bm$ or less.
The parameter $m$ is referred to as the interleaving degree.
Thus, the following theorem holds for classical burst-error-correcting codes~\cite{rf:CEC}.
\begin{th}
If there is an $[n,k]$ classical code capable of correcting all bursts of length $b$ or less, then interleaving  this code  to $m$ degree produces an $[nm,km]$ classical code with burst-error-correcting ability $bm$.
\end {th}

The $quantum$ burst-error-correcting codes are defined naturally.
Consider a set of quantum errors such that both bit and phase errors are bursts of length $b$ or less.
Then any quantum code that can correct such bursts is called a quantum burst-error-correcting code with burst-error collecting ability $b$.
An example of the bit-burst error with length $l$ is shown in Fig. 2.
In this figure, $X_* \in \left\{ X,I \right\}$.

%
%
%
%
%
%
%
%
%
%
%
\begin{center}
\begin{figure}[t]
\hspace{0.3cm}
\epsfxsize=8.5cm
\epsfbox{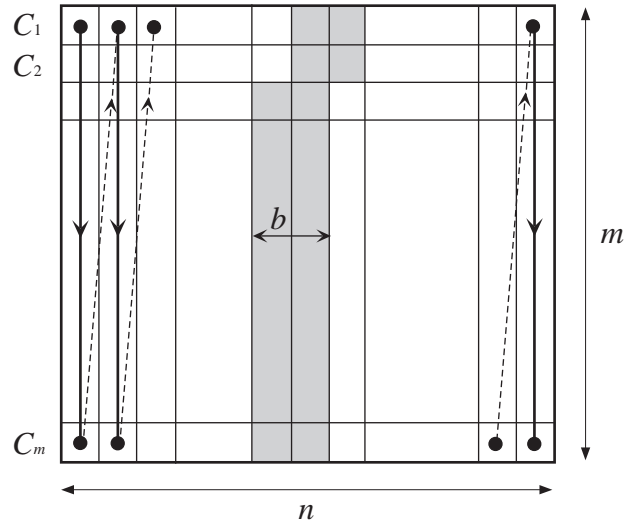}
\caption{
Construction of classical interleaved code. 
Shaded region is an example of a burst error with length $bm$.}
\end{figure}
\end{center}
%
%
%
%
%
%
%
%
\begin{center}
\begin{figure}
\hspace{0.3cm}
\epsfxsize=8.5cm
\epsfbox{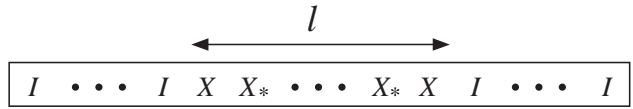}
\caption{
Example of the quantum bit-burst error with length $l$, where $X_{*} \in \left\{ X,I \right\}$.
}
\end{figure}
\end{center}
%
%
%
%
%
%
%
%
\begin{center}
\begin{figure}
\hspace{0.3cm}
\epsfxsize=8.5cm
\epsfbox{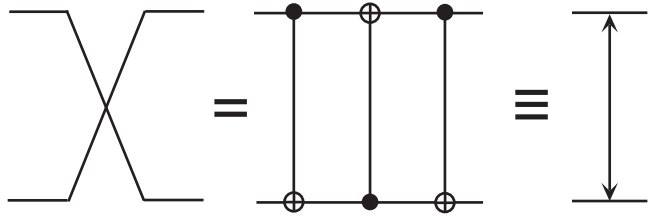}
\caption{
Quantum circuit of the swapping operation which exchanges any two quantum states.}
\end{figure}
\end{center}
%
%
%
%
%
%
%
%
\begin{center}
\begin{figure}
\hspace{0.3cm}
\epsfxsize=8.5cm
\epsfbox{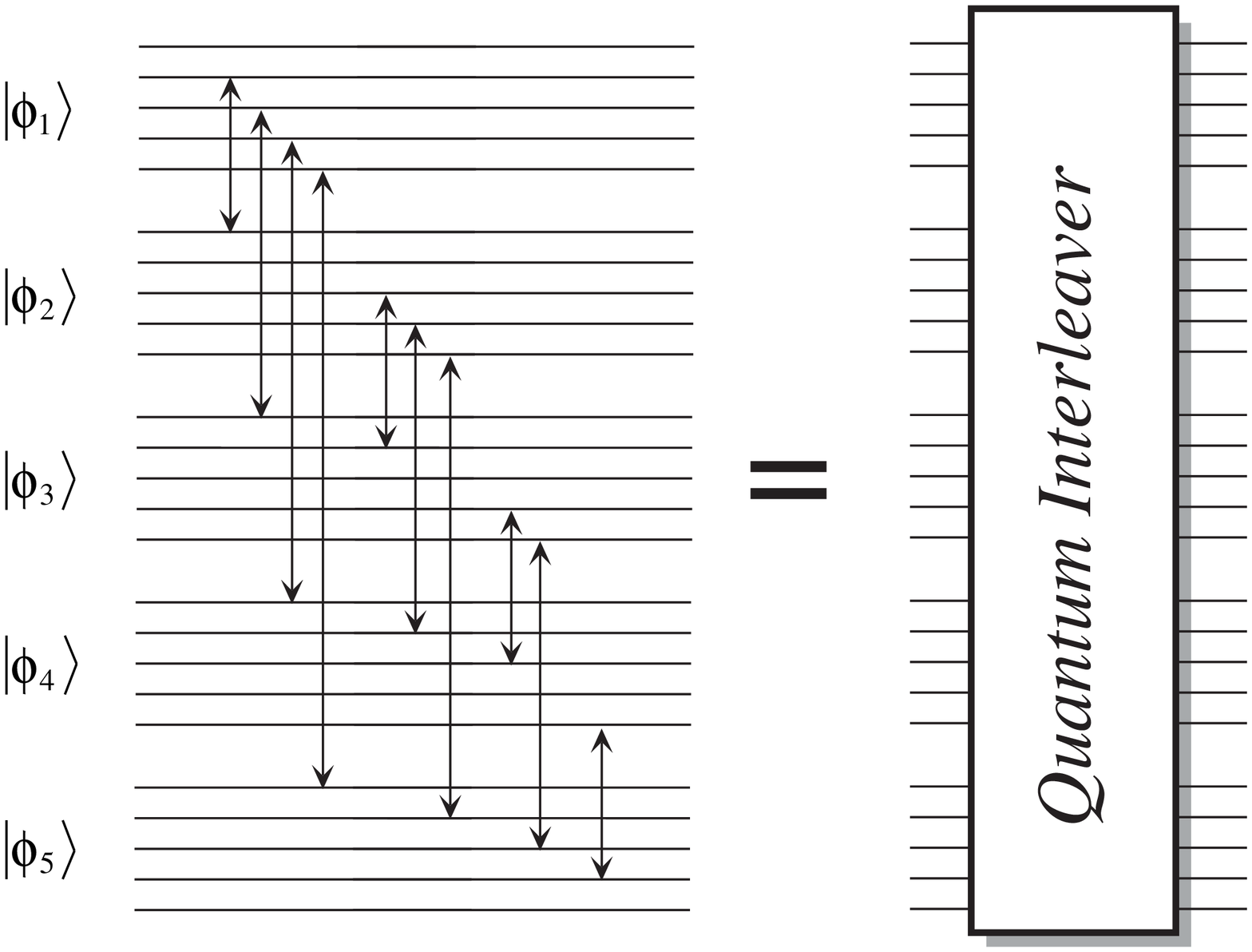}
\caption{
Quantum circuit for the quantum interleaver for the case of $n=5$ and $m=5$.}
\end{figure}
\end{center}
%
%
%
%

We show how the quantum interleaver can be transformed into quantum unitary gates.
In order to interleave the quantum codes, we must exchange the qubits one by one, following eq.~(\ref{eqn:e4}).
Therefore, the basic step of the quantum interleaving is a swapping operation for two qubits.
The swap operation $U_{swap}$ between any two quantum states can be done by using an array of three controlled not (CNOT) gates, as shown in Fig. 3.
Thus an interleaving operation $U_{int}$ can be written as tensor products of $U_{swap}$.
In the case of $n=m$, the interleaving operation is achieved simply by swapping a pair of qubits $i$ and $j$ for all $i$ and  $j(\ne i)$ as $(i ,j)\leftrightarrow(j,i)$.
Therefore, in this case, the resulting gate array consists of
\begin{eqnarray}
 3 \sum_{l=1}^{n-1} l = \frac{3}{2} n \left( n-1\right) \sim O(n^{2})
\end{eqnarray}
operations.
In Fig. 4 the circuit of the quantum interleaver is presented for the case of $n=m=5$.
In the case of $n \ne m$, the number of quantum gates is given by $O(n m)$.~\cite{rf:Kawabata}

Here we show a simple example of the quantum burst-error correction in order to analyze in detail the working of the quantum interleaver.
 Let us consider the case in which phase errors ($Z_\beta$) occur but bit errors ($X_\alpha$) never appear.
This is the simplest nontrivial case and is of practical interest.
 The simplest example of the phase-error-correcting code is that using three qubits:
\begin{eqnarray}
\left| 0 \right> \to \left| C_0 \right>
&=& \left| 000 \right> + \left| 011 \right>  +\left| 101 \right> +\left| 110 \right>
, 
\label{eqn:e6}
\\
\left| 1 \right> \to \left| C_1 \right>
&=& \left| 111 \right> + \left| 100 \right>  +\left| 010 \right> +\left| 001 \right> 
,
\label{eqn:e7}
\end{eqnarray}
which protects only against a single phase error~\cite{rf:Steane}.
Here and in the following we omit the normalization factors.
Thus, a single qubit state $c^0 \left|0\right> + c^1 \left| 1 \right>$ is encoded as $\left| \phi \right>= c^0 \left| C_0 \right> + c^1 \left| C_1 \right> $. Note that the burst $phase$-error-correcting ability of this  three qubits code is 1.

In the first step, we  prepare $m=3$ quantum code words,
\begin{eqnarray}
\left| \phi_1 \right>_{1,2,3} &=&c^0_{1} \left| C_0 \right>_{1,2,3} + c^1_{1} \left| C_1 \right>_{1,2,3}
,
\label{eqn:e8}
\\
\left| \phi_2 \right>_{4,5,6} &=& c^0_{2} \left| C_0 \right>_{4,5,6} + c^1_{2} \left| C_1 \right>_{4,5,6}
,
\label{eqn:e9}
\\
\left| \phi_3 \right>_{7,8,9}&=&c^0_{3} \left| C_0 \right>_{7,8,9} + c^1_{3} \left| C_1 \right>_{7,8,9}
,
\label{eqn:e10}
\end{eqnarray}
where the subscripts of $\left| \phi_i \right>(i=1,2,3)$, $\left| C_{0} \right>$ and $\left| C_{1} \right>$ indicate the positions of the qubits.
By interleaving these quantum codes to the $m=3$ degree, we will obtain the quantum code with $n=9$ and the burst phase-error-correcting ability 3.
In order to confirm this, we describe the nine qubits code as 
\begin{eqnarray}
\left| \phi_{in} \right>
\equiv
\left| \phi_1 \right>_{1,2,3} \otimes
\left| \phi_2 \right>_{4,5,6} \otimes
\left| \phi_3 \right>_{7,8,9}
=
\sum_{i,j,k=0}^{1}
\gamma_{i,j,k}
\left| C_{i} \right>_{1,2,3} \otimes
\left| C_{j} \right>_{4,5,6}  \otimes
\left| C_{k} \right>_{7,8,9}  
,
\label{eqn:e11}
\end{eqnarray}
where $\gamma_{i,j,k} \equiv c^i_{1} c^j_{2} c^k_{3}$.
Next we apply the interleaving operator $U_{int}$ to this code as follows:
\begin{eqnarray}
U_{int} \left| \phi_{in} \right>
=
\sum_{i,j,k=0}^{1}
\gamma_{i,j,k}
\left| C_{i} \right>_{1,4,7} \otimes
\left| C_{j} \right>_{2,5,8}  \otimes
\left| C_{k} \right>_{3,6,9}  
\equiv
\left| \phi' \right>
.
\label{eqn:e12}
\end{eqnarray}
As a result of  the interaction with the environment $\left| e \right>$, the interleaved code $\left| \phi' \right>$ undergoes the following entanglement (see eq.~(\ref{eqn:e3})):
\begin{eqnarray}
\left| \phi' \right> \otimes \left| e \right>
\to
U_{qe} \left| \phi' \right> \otimes \left| e \right>
=
\sum_{\beta}
Z_{\beta}
\left| \phi' \right> 
\otimes 
\left| e_{\beta} \right>
.
\label{eqn:e13}
\end{eqnarray}
Below, we consider the following superposition of two types of burst error with length $3$ as an example:
\begin{eqnarray}
\sum_{\beta}
Z_{\beta}
\left|  e_{\beta} \right>
&=&
Z_{111000000}
\left|  e_{111000000} \right>
+
Z_{000001110}
\left|  e_{000001110} \right>
\nonumber\\
&=&
Z_{1} \otimes Z_{2} \otimes Z_{3} \otimes I_{4} \otimes \cdots \otimes I_{9}
\left|  e' \right>
+
I_{1} \otimes \cdots \otimes I_{5} \otimes Z_{6} \otimes Z_{7} \otimes Z_{8} \otimes I_{9} 
\left|  e"\right>
,
\label{eqn:e14}
\end{eqnarray}
where $\left|  e' \right> \equiv \left|  e_{111000000} \right> $ and $\left|  e" \right>\equiv  \left|  e_{000001110} \right>$.
The resulting state is given by
\begin{eqnarray}
U_{qe} \left| \phi' \right> \otimes \left| e \right>
=
\sum_{i,j,k=0}^{1}
\gamma_{i,j,k}
\left[
Z_1 \left| C_{i} \right>_{1,4,7} \otimes
Z_2 \left| C_{j} \right>_{2,5,8}  \otimes
Z_3 \left| C_{k} \right>_{3,6,9}  
\otimes 
\left|  e' \right>
\right.
\nonumber\\
+
\left.
Z_7 \left| C_{i} \right>_{1,4,7} \otimes
Z_8 \left| C_{j} \right>_{2,5,8}  \otimes
Z_6 \left| C_{k} \right>_{3,6,9}  
\otimes 
\left|  e" \right>
\right]
.
\label{eqn:e15}
\end{eqnarray}
Here we have omitted identity operators $I$ for simplicity.
Finally we apply the operator $U_{int}$ in order to deinterleave the interleaved quantum code:
\begin{eqnarray}
&&
U_{qe} \left| \phi' \right> \otimes \left| e \right>
\nonumber\\
&\to&
U_{int} U_{qe} \left| \phi' \right> \otimes \left| e \right>
\nonumber\\
&=&
\sum_{i,j,k=0}^{1}
\gamma_{i,j,k}
\left[
Z_1 \left| C_{i} \right>_{1,2,3} \otimes
Z_4 \left| C_{j} \right>_{4,5,6}  \otimes
Z_7 \left| C_{k} \right>_{7,8,9}  
\otimes 
\left|  e' \right>
+
Z_3 \left| C_{i} \right>_{1,2,3} \otimes
Z_6 \left| C_{j} \right>_{4,5,6}  \otimes
Z_8 \left| C_{k} \right>_{7,8,9}  
\otimes 
\left|  e" \right>
\right]
\nonumber\\
&=&
Z_{1} \left| \phi_1 \right>_{1,2,3}
\otimes
Z_{4} \left| \phi_2 \right>_{4,5,6}
\otimes
Z_{7} \left| \phi_2 \right>_{7,8,9}
\otimes
\left|  e' \right>
+
Z_{3} \left| \phi_1 \right>_{1,2,3}
\otimes
Z_{6} \left| \phi_2 \right>_{4,5,6}
\otimes
Z_{8} \left| \phi_2 \right>_{7,8,9}
\otimes
\left|  e"  \right>
,
\label{eqn:e16}
\end{eqnarray}
where we utilize the fact that $U_{int}$ does not operate on the environment.
Therefore, ($b=3$) burst phase errors  in the code $\left| \phi_{in} \right>$ now appear as ($b=1$) short errors in each code word $\left| \phi_{i} \right>$ ($i=1,2,3$).
These short errors are correctable independently in each three qubit code word by following the procedure described in  ref. 9.

The extension of this analysis to the case of general errors is straightforward and will be given elsewhere~\cite{rf:Kawabata}.
If we interleave the [[5,1]] quantum code~\cite{rf:5qubits1,rf:5qubits2,rf:5qubits3} with burst-error-correcting ability $b=1$ to the $m=5$ degree, then we can obtain the $[[25,5]]$ quantum code with burst-error-correcting ability $bm=5$.
 Here, we have used the notation $[[n,k]]$ to refer to a quantum error-correcting code for $n$ qubits having $2^k$ code words according to ref. 12.

The following theorem shows that a long quantum code with good burst-error-correcting properties can be constructed by interleaving a short quantum code.
\begin{th}
If there is an $[[n,k]]$ quantum code capable of correcting all bursts of length $b$ or less, then interleaving  this code  to $m$ degree produces an $[[nm,km]]$ quantum code with burst-error-correcting ability $bm$.
\end {th}

The general networks for the encoding and decoding procedures of the $[[n,k]]$ quantum codes which produce the $[[nm,km]]$ quantum code are shown in Fig. 5.

In summary, we have proposed a simple method for constructing quantum burst-error-correcting cods without increasing redundant qubits.
By using the quantum interleaving method, the quantum code words can be distributed amongst the qubit stream so that consecutive words are never next to each other.
On deinterleaving they are returned to their original positions so that any errors that have occurred become widespread. 
This ensures that any burst (long) errors now appear as random (short) errors.
Finally, it should be pointed that the classical interleaver is currently widely used, not only in error correction but also in various digital communications and devices.
We hope that the concept of the quantum interleaver proposed in this letter may ultimately have a variety of applications for quantum communication, quantum computation, and quantum engineering.

The author wishes to thank S. Abe,  N. Imoto and M. Koashi for valuable discussions and comments.

%
%
%
%
\begin{center}
\begin{figure}
\hspace{0.3cm}
\epsfxsize=8.5cm
\epsfbox{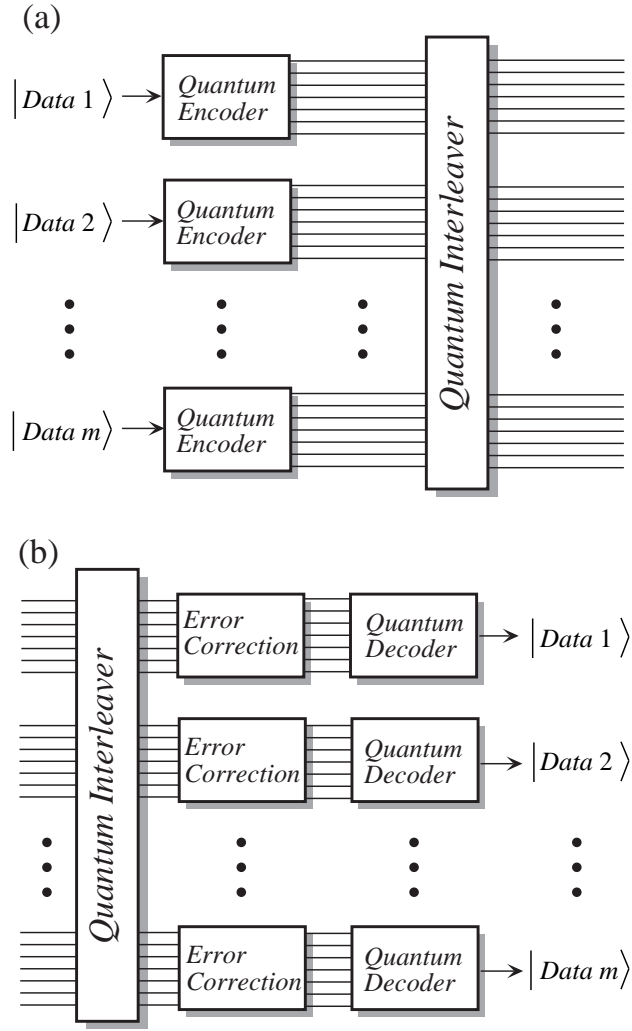}
\caption{
Quantum network for (a) encoding, and (b) error-correcting and decoding procedure.}
\end{figure}
\end{center}
%
%
%
%

%
%
%
%

%
\end{document}